\documentclass[twocolumn,aps,amsmath,amssymb,floatfix]{revtex4}

\usepackage{graphicx}
\usepackage{dcolumn}
\usepackage{bm}
\usepackage{setspace}
\usepackage{color}

\begin{document}

\title{Polar Chemoreceptor Clustering by Coupled Trimers of Dimers}

\author{Robert G.~Endres}
\affiliation{
Division of Molecular Biosciences {\it and} 
Centre for Integrated Systems Biology at Imperial College, 
Imperial College London, London SW7 2AZ, United Kingdom
}

\date{\today}

\begin{abstract}
Receptors of bacterial chemotaxis form clusters at the cell poles, where clusters act as ``antennas'' 
to amplify small changes in ligand concentration. 
Interestingly, chemoreceptors cluster at multiple length scales.
At the smallest scale, receptors form dimers, which assemble into stable
timers of dimers. At a large scale, trimers form large polar clusters composed of
thousands of receptors.
Although much is known about the signaling properties emerging from receptor clusters, it is unknown 
how receptors localize at the cell poles and what the cluster-size determining factors are. Here, we present a model 
of polar receptor clustering based on coupled trimers of dimers,
where cluster size is determined as a minimum of the cluster-membrane free energy. This energy has contributions
from the cluster-membrane elastic energy, penalizing large clusters due to their high intrinsic curvature, 
and receptor-receptor coupling favoring large clusters. 
We find that the reduced cluster-membrane curvature mismatch at the curved cell poles
leads to large and robust polar clusters in line with experimental 
observation, while lateral clusters are efficiently suppressed. 
\end{abstract}

\maketitle

\section{Introduction}

Chemoreceptor clustering is widely conserved among bacteria and archaea \cite{gestwicki00}, 
allowing cells to detect chemicals 
in the environment with high sensitivity over a wide range of background concentrations. In the bacteria 
{\it Escherichia coli}, {\it Salmonella enterica}, and {\it Caulobacter crescentus}, receptor
clustering is well documented and occurs at multiple length scales. At a small scale, 
chemotaxis receptors form stable homodimers, which then assemble into larger complexes with 
receptors of different chemical specificities intermixed \cite{studdert02}. Three homodimers,
connected at their signaling tip, form a trimer of dimers (named {\it trimer} from here on) 
\cite{kim99,ames02,studdert02}, believed to be the smallest stable signaling unit \cite{studdert05,boldog06}. 
At a larger scale, thousands of receptors 
\cite{li04} form approximately 200-nm large polar clusters ({\it cf.} Fig. 1a) 
\cite{maddock93,sourjik00,ames02,thiem07,zhang07,briegel08}. Despite the excellent characterization of 
much of the bacterial chemotaxis network, it is unknown how receptors localize at the cell poles 
and how they assemble into large polar clusters.

Polar localization appears to be an intrinsic property of chemoreceptors \cite{kentner06b,schulmeister08}. 
It hardly depends on the presence or absence of the receptor-bound kinase CheA and adapter protein CheW \cite{kentner06a}, 
and is unaffected by removal of the periplasmic ligand-binding domain of the receptors \cite{kentner06a}. 
It is also a passive process since newly synthesized receptors, 
initially inserted at random positions in the membrane, diffuse and ultimately 
become trapped at the cell poles \cite{shiomi06,yu06}. 
Most importantly, polar localization appears to depend on membrane curvature. 
First, inhibition of actin-homologue MreB 
in growing cells leads to cell swelling and a diffuse receptor distribution, 
with remaining receptor localization in areas of increased cell curvature \cite{shih05}.  
Second, receptor-membrane extracts self-assemble into round micelles after receptor 
overexpression and cell lysis \cite{weis03} (Fig. 1b). From electron micrographs, the 
intrinsic curvature of the trimer structure can be estimated \cite{lefman04}. 
Third, other two-component receptor dimers, {\it e.g.} the receptor LuxQ of the quorum-sensing pathway, 
dimerize without forming trimers of dimers and are evenly distributed over the 
cell surface \cite{fred06}. Taken together, these observations suggest that the distinct 
trimer structure with its increased intrinsic curvature is responsible for polar receptor 
localization.

While trimers may have a tendency to localize at the cell poles and areas of high membrane curvature, 
tight clustering requires an attractive coupling among the trimers. The conventional 
view is that CheA and CheW mediate interactions among receptors. 
Alternative models include swapping of the cytoplasmic receptor 
domains \cite{wolanin04} and membrane-mediated coupling \cite{ursell07} (see Results and Discussion section). 
The high sensitivity and cooperativity obtained 
from {\it in vivo} FRET (fluorescence resonance energy transfer) 
\cite{sourjik02a,sourjik04} and {\it in vitro} 
\cite{weis00} data demonstrate that the functional units of receptor signaling 
are indeed larger than trimers. These observations are supported by recent 
quantitative models \cite{bray98,sourjik04,tu05,keymer06,endres06,skoge06}.

\begin{figure}            
\includegraphics[width=8cm,angle=0]{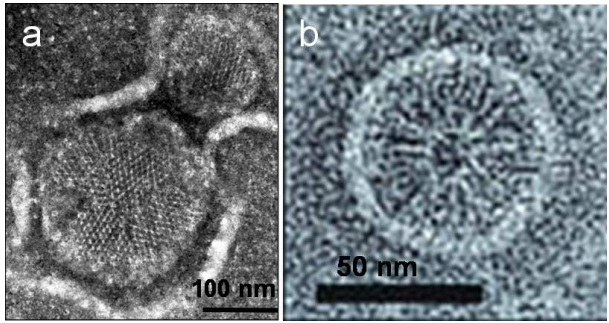}
\caption{Electron micrographs of chemoreceptor clusters. 
(a) Extended clusters in membrane preparations \cite{mcandrew04,lai05,mcandrew05,mcandrew06}. Clusters of similar size 
are observed at the poles of living cells \cite{zhang07,briegel08}. Image is courtesy of Michael Manson. 
(b) Self-assembled round micelle at receptor-dimer resolution \cite{weis03}.
Image printed with written permission from publisher.
All experiments are based on Tsr-receptor overexpression, as well as membrane extraction, 
negative staining, and freezing for imaging.}
\label{fig:fig1}
\end{figure}

Based on these observations, we propose a model for polar receptor localization and clustering 
due to the high intrinsic curvature of trimers and an attractive trimer-trimer coupling. 
Specifically, we consider a membrane-embedded cluster composed of trimers. For a
sphero-cylindrical cell, we assume that the average membrane curvature at the poles is twice as large as
average curvature at the lateral surface area, and that trimers have a high intrinsic curvature (Fig. 2a). 
The intrinsic curvature of a trimer tends to deform the membrane (Fig. 2c), 
penalizing large clusters of trimers. However, attractive coupling between trimers favors cluster formation, 
leading to a competition between these two opposing energy contributions. 
Using continuum elastic theory, we derive an analytical expression for the total cluster-membrane energy. 
We find that, due to the reduced cluster-membrane curvature mismatch at the 
poles, trimers favorably cluster at the poles and not at the lateral cell area. 
Furthermore, the cluster-size distribution is determined by the cluster-membrane energetics, 
as well as the trimer density in the cell membrane. 
Our predicted average cluster size is in line with experimental observation.

\section{Model}

\noindent{\bf Receptor geometry}\\
In our model, receptor dimers are assumed to always be associated in trimers, the smallest
stable signaling unit \cite{studdert05,boldog06}.
In Fig. 2a, the receptor dimer length and width, as well as the distance between neighboring dimers within a trimer 
are taken from partial crystal structures \cite{kim99} and electron microscopy \cite{weis03}. Very similar
parameters were used to model the physical response of trimers to osmolytes measured by homo-FRET \cite{vaknin06}.
Importantly, the estimated value for the intrinsic curvature of a trimer corresponds closely to 
the inverse radius of self-assembled micelles (Fig. 1b, \cite{weis03}).
The value for the trimer cross section $A$ (table I) is consistent with a 
three-dimensional model of the receptor cluster \cite{shimizu00,kim02} and
estimates from cryo-electron microscopy \cite{weis03,mcandrew04,briegel08}. \\

\noindent{\bf Elastic cluster-membrane energy}\\
The elastic energy of a membrane-embedded cluster is determined by
cluster and membrane bending energies and a pinning potential
\begin{eqnarray}
E_{el}&=&\int_{c} \left[\frac{\kappa_{c}}{2}(2\bar C_c(\vec r)-C_T)^2+
\frac{\lambda}{2}(\tilde h_{c}(\vec r)-h_0(\vec r))^2\right]d^2\vec r\nonumber\\
&&\!\!\!\!\!\!\!\!\!\!\!\!\!\!\!\!\!\!\!\!\!\!\!\!\!+\int_{m} \left[\frac{\kappa_{m}}{2}(2\bar C_m(\vec r)-C_m)^2+
\frac{\lambda}{2}(\tilde h_m(\vec r)-h_0(\vec r))^2\right]d^2\vec r.\label{eq:Eel0}
\end{eqnarray}
The first term, proportional to the bending stiffness $\kappa_c$ of the receptor cluster,
penalizes deviations between the total cluster curvature $2\bar C_c$ and the preferred
cluster curvature, which is equal to the intrinsic trimer curvature $C_T$. 
The total cluster curvature $2\bar C_c=C_1+C_2$ is defined by the two principal curvatures 
$C_1$ and $C_2$
 \cite{boal02}. The second term in Eq. \ref{eq:Eel0}, proportional to the 
pinning modulus $\lambda$ \cite{huang06,ranjan08}, penalizes deviations of the
cluster height $\tilde h_c$ from the preferred height $h_0$ determined by the shape of the
curved cell wall. The third and forth terms in Eq. \ref{eq:Eel0} mirror the first and second, 
respectively, and describe the cluster-surrounding membrane with total curvature $2\bar C_m$, 
preferred curvature $C_m$, and height $\tilde h_m$. Bending stiffnesses $\kappa_{c}$ and $\kappa_m$ 
arise from optimal packing of receptors and lipids, respectively, aiming to protect hydrophobic residues 
from polar water. The pinning modulus arises due to the turgor pressure, which pushes the membrane and cluster
outward, while the rigid cell wall pushes them inward. The net effect is a penalty for deformations 
away from the preferred cell shape (see Fig. 2) \cite{huang06,ranjan08}.

\begin{figure}
\includegraphics[width=8cm,angle=0]{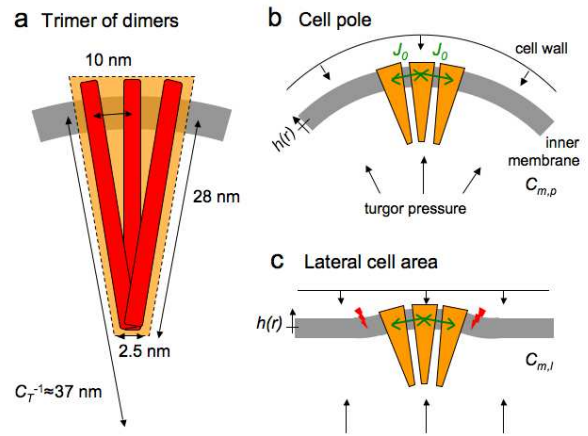}
\caption{Schematic of membrane-inserted receptors. (a) Trimer (orange shaded area) 
of dimers (red bars) with geometric parameters. $C_T$ is the intrinsic curvature of a trimer.
(b) Cluster of three trimers at the cell pole. Trimer-trimer coupling strength $J_0$ 
is indicated by green arrows. Also shown are the cell wall and the inner membrane (gray) 
with curvature $C_{m,p}$. The height profile $h(r)$ describes the cluster-membrane deformation
as measured relative to the preferred height due to cell wall and turgor pressure. 
(c) Same cluster at a lateral position with membrane curvature $C_{m,l}$.
Red arrows indicate energetically unfavorable membrane deformations due to the
cluster-membrane curvature mismatch.}
\label{fig:fig2}
\end{figure}

Let us define the relative height perturbation $h_{c(m)}=\tilde h_{c(m)}(\vec r)-h_0(\vec r)$
as measured relative to the preferred height $h_0(\vec r)$ in the direction of the normal, pointing in
the radial direction outward from surface $h_0(\vec r)$ (Fig. 2 b,c). This allows us to
perform the following calculations in the so-called normal gauge, where the total curvatures for the
cluster (and membrane) $2\bar C_{c(m)}$ are now given by $C_{m}+(C_m^2-2C_G)h_{c(m)}+\nabla^2h_{c(m)}$
to lowest order, following from the first-order variation of the geometry \cite{capovilla03}.
Here, $C_G$ is the Gaussian curvature. We use two further approximations: (1)  
$\nabla^2h_{c(m)}>\!\!>(C_m^2-2C_G)h_{c(m)}$, justified for small amplitude and large bending ripples.
Hence, the contribution proportional to $h_{c(m)}$ is neglected. (2) The curvilinear Laplacian 
is replaced by the flat Laplacian $\nabla^2\approx\partial^2/\partial x^2+\partial^2/\partial y^2$, 
valid for sufficiently small clusters.
Introducing $2\bar C_{c(m)}\approx C_m+\nabla^2h_{c(m)}$ and $\Delta C=C_T-C_m$, Eq. \ref{eq:Eel0} 
can thus be written as
\begin{widetext}
\begin{equation}
E_{el}=\int_{c} \left[\frac{\kappa_{c}}{2}(\nabla^2h_{c}(\vec r)-\Delta C)^2+
\frac{\lambda}{2}h_{c}^2(\vec r)\right]d^2\vec r+\int_{m} \left[\frac{\kappa_{m}}{2}(\nabla^2h_m(\vec r))^2+
\frac{\lambda}{2}h_m^2(\vec r)\right]d^2\vec r.\label{eq:Eel}
\end{equation}
\end{widetext}
Parameter $\Delta C$ is the important {\it curvature mismatch}, {\it i.e.} the difference 
between the cluster and membrane curvatures.
The elastic energy model in Eq. \ref{eq:Eel} neglects surface tension, as well as an
Gaussian curvature effect due to the different cluster and membrane elastic 
properties (see Appendix A for a justification). In this paper, the model is applied to both
polar and lateral clusters using standard parameter values given in table \ref{table1}.
Very similar models were previously applied to describe mixtures of lipids with different 
curvatures \cite{lipowsky92,seifert93,komura06,huang06,ranjan08}. 
For a general review and alternative elastic energy models, see \cite{kozlov96,brown08}.

Considering a circular cluster of radius $R$, the total height profile $h(\vec r)=h(x,y)$ is composed of 
$h_c(\vec r)$ for the cluster 
($r<R$) and $h_m(\vec r)$ for the membrane ($r>R$), and is determined by minimizing the total elastic 
cluster-membrane energy with respect to variation of $h(x,y)$. Following Ref. \cite{nielson98}, 
minimizing an elastic energy of the generic form
\begin{equation}
E_{el}^g=\int_a^b\int_c^d\Psi\left(x,y,h,\frac{\partial h}{\partial x},\frac{\partial h}{\partial y},
\frac{\partial^2 h}{\partial x^2},
\frac{\partial^2 h}{\partial y^2}\right)dx\,dy
\end{equation}
leads to the following Euler-Lagrange equation
\begin{eqnarray}
\frac{\partial \Psi}{\partial h}&-&\frac{\partial}{\partial x}\left(\frac{\partial\Psi}{\partial(\partial h/\partial x)}\right)
-\frac{\partial}{\partial y}\left(\frac{\partial\Psi}{\partial(\partial h/\partial y)}\right)\\
&&\!\!\!\!\!\!\!\!\!\!\!\!\!\!\!\!\!\!\!+\frac{\partial^2}{\partial x^2}\left(\frac{\partial\Psi}
{\partial(\partial^2h/\partial x^2)}\right)+
\frac{\partial^2}{\partial y^2}\left(\frac{\partial\Psi}{\partial(\partial^2h/\partial y^2)}\right)=0.
\end{eqnarray}
Replacing $\Psi$ by the expressions in Eq. \ref{eq:Eel} leads to the forth-order linear differential equation 
\cite{huang86,dan98,wiggins05} for cluster (membrane)
\begin{equation}
\nabla^4h_{c(m)}(\vec r)+\frac{\lambda}{\kappa_{c(m)}} h_{c(m)}(\vec r)=0\label{eq:ELcm}
\end{equation}
which is independent of $\Delta C$ in the small deformation approximation. 
Since we consider a circular cluster located at the origin of 
the coordinate system, we apply cylindrical symmetry from now on.
To solve Eq. \ref{eq:ELcm} for cluster and membrane one needs four boundary conditions for each equation. We require
$\partial h_c/\partial r=0$ at the origin and that the membrane deformation vanishes far away from the cluster,
{\it i.e.} $\lim_{r\rightarrow \infty}h_m(r)=0$ and $\lim_{r\rightarrow \infty}\partial h_m(r)/\partial r=0$. 
We further impose that the solutions for cluster and membrane 
match at the cluster-membrane interface, {\it i.e.} $h_c(R)=h_m(R)$ and 
$\partial h_c/\partial r|_{R}=\partial h_m/\partial r|_{R}$. 

Eq. \ref{eq:ELcm} can be solved by applying the Kelvin differential equation
\begin{equation}
\nabla^2h(\beta r)-i\beta^2h(\beta r)=0\label{eq:Kelvin}
\end{equation}
leading to $\beta_{c(m)}=\sqrt[4]{\frac{\lambda_{c(m)}}{\kappa_{c(m)}}}$ for the cluster (membrane).
In Eq. \ref{eq:Kelvin}, the second-order derivative is now calculated 
using $\nabla^2=\partial^2/\partial r^2+1/r\,\partial/\partial r$.
The solution to Eq. \ref{eq:Kelvin} is given by \cite{AS68}
\begin{eqnarray}
&&ber_0(\beta_{c(m)}r)+ibei_0(\beta_{c(m)}r)=I_0(\beta_{c(m)}re^{-i3\pi/4})\\
&&ker_0(\beta_{c(m)}r)+ikei_0(\beta_{c(m)}r)=K_0(\beta_{c(m)}re^{i\pi/4})
\end{eqnarray}
where $ber_0, bei_0, ker_0$, and $kei_0$ are the zeroth-order Kelvin function, and $I_0$ and 
$K_0$ are the zeroth-order modified Bessel functions of the first and second kind, respectively.

To construct the solution for the cluster ($r<R$), only the modified Bessel function of the first
kind has zero slope at $r=0$, whereas to construct the solution for the membrane ($r>R$) only the 
modified Bessel function of the second kind has a vanishing real part at infinity. To obtain real 
solutions, we need to add their complex conjugates
\begin{eqnarray}
\!\!\!\!\!\!\!\!\!\!\!h_c(r)\!&=&\!(a+ib)I_0(\beta r e^{-i3\pi/4})\!+\!(a-ib)I_0(\beta r e^{+i3\pi/4})\label{eq:solc}\\
\!\!\!\!\!\!\!\!\!\!\!h_m(r)\!&=&\!(c+id)K_0(\beta r e^{+i\pi/4})\!+\!(c-id)K_0(\beta r e^{-i\pi/4})\label{eq:solm}
\end{eqnarray}
where $a, b, c$, and $d$ are real parameters to be determined by matching the boundary conditions.\\

\noindent{\bf Solution coefficients}\\
After matching boundary conditions, the coefficients of the solution 
in Eqs. \ref{eq:solc} and \ref{eq:solm} are given by
\begin{eqnarray}
a&=&\frac{H_0}{2}\label{eq:a}\\
b&=&\frac{H_0\Re I_0(\beta_{c,2}R)-H_1}{2\Im I_0(\beta_{c,2}R)}\label{eq:b}\\
c&=&\frac{H_1+2d\Im K_0(\beta_{m,1}R)}{2\Re K_0(\beta_{m,1}R)}\label{eq:c}\\
d&=&\frac{S\Re K_0(\beta_{m,1}R)+H_1\Re[\beta_{m,1}K_1(\beta_{m,1}R)]}{2\Im[\beta_{m,1}K_0(\beta_{m,4})K_1(\beta_{m,1}R)]}\label{eq:d},
\end{eqnarray}
where $S=2\Re[(a+ib)\beta_{c,2}I_1(\beta_{c,2}R)]$ is the slope at the cluster-membrane interface, 
$H_0$ is the cluster height at $r=0$, and $H_1$ is the height at $r=R$, {\it i.e.} at the cluster-membrane interface. 
We further used $\partial I_0(\beta r)/\partial (\beta r)=\beta I_1(\beta r)$ 
and $\partial K_0(\beta r)/\partial (\beta r)=-\beta K_1(\beta r)$.
Remaining unknown parameters, such as $H_0$ and $H_1$, are determined by numerically minimizing the elastic energy.\\

\noindent{\bf Analytic expression for elastic energy}\\
The integrals in Eq. \ref{eq:Eel} can be solved analytically using integration-by-parts twice 
\cite{landau64,wiggins05} and exploiting Eq. \ref{eq:ELcm}, leading to
\begin{widetext}
\begin{equation}
E_{el}=\pi\kappa_cR\{S\cdot\nabla^2h_c|_R-H_1\cdot\nabla^3 h_c|_R-2\Delta C\cdot S\}+\frac{1}{2}\pi\kappa_c\Delta C^2R^2-\pi\kappa_mR\{S\cdot\nabla^2h_m|_R-H_1\cdot\nabla^3 h_m|_R\}\label{eq:Eel2}.
\end{equation}
\end{widetext}
The higher-order derivatives are calculated from the solution Eqs. \ref{eq:solc} and \ref{eq:solm} using \cite{AS68}
\begin{eqnarray}
\nabla I_0(\beta r)&=&\frac{\partial I_0}{\partial r}=\beta I_1(\beta r)\\
\nabla^2I_0(\beta r)&=&\left(\frac{\partial^2}{\partial r^2}+\frac{1}{r}\frac{\partial}{\partial r}\right)I_0(\beta r)\nonumber\\
&=&\beta^2I_0(\beta r)\\
\nabla^3I_0(\beta r)&=&\frac{\partial}{\partial r}\nabla^2I_0(\beta r)\nonumber\\
&=&\left(\frac{\partial^3}{\partial r^3}+\frac{1}{r}\frac{\partial^2}{\partial r^2}-
\frac{1}{r^2}\frac{\partial}{\partial r}\right)I_0(\beta r)\nonumber\\
&=&\beta^3I_1(\beta r)
\end{eqnarray}
and 
\begin{eqnarray}
\nabla K_0(\beta r)&=&-\beta K_1(\beta r)\\
\nabla^2K_0(\beta r)&=&\beta^2K_0(\beta r)\\
\nabla^3K_0(\beta r)&=&-\beta^3K_1(\beta r).
\end{eqnarray}

In the limit of very large clusters, the elastic energy Eq. \ref{eq:Eel2} reduces to
\begin{equation}
E_{el}^{(\infty)}\longrightarrow \frac{1}{2}\pi\kappa_c\Delta C^2R^2=\frac{1}{2}\kappa_c\Delta C^2AN,
\end{equation}
since the other contributions to the elastic energy grow more slowly with size ({\it cf.} Fig. 5a).
The number of trimers in a cluster of radius $R$ is given by $N\approx R^2\pi/A$.\\

\noindent{\bf Attractive trimer-trimer coupling}\\
In our model, trimers interact favorably when in close contact, 
driving cluster formation. The total coupling energy of a cluster of radius $R$ 
is described by
\begin{equation}
E_\text{a}=-J(N)\cdot N.\label{eq:Ea}
\end{equation}
For a triangular lattice used in Fig. 3, we consider the following two 
expressions for the average coupling energy per trimer.
Assuming a cluster made of concentric rings formed around a central trimer, the
coupling energy is given by \cite{ranjan08}
\begin{equation}
J(N)=J_0\left(3-\sqrt{\frac{12}{N}-\frac{3}{N^2}}\right)\label{eq:JN1},
\end{equation} 
where $J_0$ is the coupling energy between two neighboring trimers (table I).
This expression is exact in the limit of large clusters. Alternatively, we fit 
\begin{equation}
J(N)=3J_0\frac{N}{N+N_{0.5}}\label{eq:JN2}.
\end{equation} 
to exact interaction energies of small compact clusters, and obtain
parameter $N_{0.5}=4.72$. According to Fig. 3, Eq. \ref{eq:JN1} overestimates the 
interaction energy for
small clusters, whereas Eq. \ref{eq:JN2} overestimates the interaction energy
for large clusters. In the limit $N\rightarrow\infty$, both models produce 
$E_\text{a}^{(\infty)}\rightarrow -3J_0N$. In the following we use Eq. \ref{eq:JN2} 
since it helps stabilizing large clusters.

The total energy is the sum of elastic energy and the attractive energy
\begin{equation}
E(N)=E_{el}(N)+E_a(N),\label{eq:Etot}
\end{equation} 
where we explicitly included the dependence on the cluster size $N$.\\

\noindent{\bf Distribution of cluster sizes}\\
For a sphero-cylindrical cell, the preferred membrane curvature at the 
poles $C_m=C_{m,p}$ is twice as large as the preferred membrane curvature at the 
lateral area $C_m=C_{m,l}=C_{m,p}/2$, however smaller than the trimer curvature $C_T$. 
This leads to a smaller cluster-membrane curvature mismatch at the poles $\Delta C_p=C_T-C_{m,p}$
than at the lateral area $\Delta C_l=C_T-C_{m,l}=C_T-C_{m,p}/2$. 
Consequently, the total energy Eq. \ref{eq:Etot} of a cluster is also
smaller at the poles than at the lateral area $E_{p}(N)<E_{l}(N)$,
favoring polar clustering.

Based on the total energies, statistical mechanics is used to calculate
the cluster-size distribution at the poles/lateral area ($p/l$) \cite{endres07,ranjan08}
\begin{equation}
P_{p/l}(N)=Ne^{-[E_{p/l}(N)-N\mu]/k_BT},\label{eq:P}
\end{equation}
where the exponential Boltzmann factor describes the probability to observe
a cluster of $N$ trimers at the poles/lateral area for chemical potential $\mu$. The chemical potential 
represents the energy required or released by inserting
a trimer into the membrane, and is adjusted to fulfill an overall target trimer density on the cell
surface (occupancy fraction) $\rho$ via
\begin{equation}
\sum_{N}[P_p(N)+P_l(N)]=\rho.\label{eq:r}
\end{equation}
Using the cluster-size distributions, the average cluster sizes at the poles and lateral area are given by
\begin{equation}
\langle N\rangle_{p/l}=\sum_NNP_{p/l}(N).
\end{equation}
\\

\noindent{\bf Conditions for large polar clusters}\\
To find the conditions which favor polar clustering, we consider the total energy density $\epsilon=E_{p/l}/N$, 
{\it i.e.} the total energy of the membrane-embedded cluster per trimer. 
Generally, minimization with respect to $N$ 
determines the energetically preferred cluster size. We note the following.
First, the energy density is generally a monotonically decreasing function of $N$, which eventually 
saturates for large $N$ ({\it cf.} Fig. 5a). This indicates that maximal cluster sizes are energetically
favorable.
Second, the elastic energy density is always smaller at the poles than at the lateral area,
demonstrating that polar clustering is energetically predominant.

Let us consider the total energy densities in the limit $N\rightarrow\infty$. In this limit, 
the total cluster-membrane energy density at the poles/lateral area is given by 
\begin{eqnarray}
\epsilon_{p/l}^{(\infty)}&=&\left(\frac{1}{2}\kappa_c\Delta C_{p/l}^2A-3J_0\right).\label{eq:Etotinf}
\end{eqnarray}
Consequently, the energy-density difference between the poles and lateral area is provided by
\begin{eqnarray}
\Delta \epsilon&=&\epsilon_{l}^{(\infty)}-\epsilon_{p}^{(\infty)}\nonumber\\
&=&\frac{\kappa_cA}{2}(C_T-3/4\,C_{m,p})C_{m,p},\label{eq:dE}
\end{eqnarray}
where we used $C_{m,l}=C_{m,p}/2$. Increasing $C_T$ beyond $3/4\,C_{m,p}$ favors polar over
lateral clusters. Specifically, for $N\Delta \epsilon>1k_BT$, a cluster of $N$ trimers 
is significantly more favorable at the poles than at the lateral area at temperature $T$.\\

\begin{table}[t]
\caption{{\bf Summary of standard parameters.} These parameters are used throughout calculations
unless specified otherwise.}
\begin{tabular}{c|c|l}Parameter                      &  Value       &  Meaning \\ 
\hline
$\lambda$ [$k_BT/\text{nm}^4$] &  $0.25$ \cite{huang06}     &  Pinning modulus\\
$\kappa_c$ [$k_BT$]            &  $120$  ${}^*\ \ \ $     &  Bending stiffness of cluster\\
$\kappa_m$ [$k_BT$]            &  $25$   \cite{huang06}     &  Bending stiffness of membrane\\
$C^{-1}_{m,p}$ [nm]            &   400   \cite{boal02}      &  Inverse of polar membrane curvature\\
$C^{-1}_{m,l}$ [nm]            &   800 ${}^\dagger\ \ \ $       &  Inverse of lateral membrane curvature\\
$C^{-1}_T$   [nm]              &  37   \cite{weis03}  &  Inverse of trimer-of-dimer curvature\\
$J_0$ [$k_BT$]                 &   3  ${}^*\ \ \ $       &  Trimer-trimer coupling strength\\
$A$ [nm${}^2$]                 &  200  \cite{shimizu00}  &  Trimer cross section
\end{tabular}
\label{table1}
${}^*\ $these parameters are varied in Fig. 5C to check for robustness.
${}^\dagger\ $based on a sphero-cylindrical cell.
\end{table}

\section{Results and Discussion}
Based on experimental observations outlined in the Introduction section, we propose a model for polar receptor 
localization and clustering (Fig. 2). The ingredients and model assumptions are as follows: 
(1) An individual trimer of dimers (trimer), believed to be the smallest stable signaling unit \cite{studdert05,boldog06}, 
has a high intrinsic curvature $C_T$ (Fig. 2a). The cell membrane has a higher curvature at the cell poles 
$C_{m,p}$  than at the lateral area $C_{m,l}$. For a sphero-cylindrical cell, we have specifically 
$C_{m,l}=C_{m,p}/2$. Since $C_T>C_{m,p}>C_{m,l}$, 
individual trimers favor the cell poles energetically, although this effect is very small 
by itself (fraction of thermal energy $k_BT$). 
(2) Trimers are coupled with strength $J_0$ when in close proximity (Fig. 2bc), driving cluster formation 
at the poles and lateral area (Fig. 3). 
(3) Due to the cluster-membrane curvature mismatch, growing clusters deform the membrane and are energetically
penalized. However, since the cluster-membrane curvature mismatch at the poles $\Delta C_{p}=C_T-C_{m,p}$ is
smaller than the corresponding mismatch at the lateral area $\Delta C_{l}=C_T-C_{m,p}/2$, this
energy penalty is reduced at the poles (Fig. 2b).
As outlined in the Model section, the model is implemented by considering a membrane-embedded cluster 
of radius $R$. The height profile of the cluster and the membrane minimizes the elastic energy, which is determined
by the cluster and membrane preferred curvatures (respective $C_T$ and $C_{m,p/l}$) and their 
bending stiffnesses (respective $\kappa_c$ and $\kappa_m$). 
Furthermore, a pinning modulus $\lambda$ \cite{huang06,ranjan08} pushes the membrane and 
the cluster against the rigid cell wall (Fig. 2 bc). This penalizes 
large deformations of the cluster and the membrane. 
The main findings are as follows: (1) Considered separately, poles and lateral area 
favor maximal cluster sizes energetically. (2) Actual cluster size is determined by timer density (entropy), 
where increasing trimer density pushes distribution of cluster sizes to larger values. 
(3) Polar-only clustering is a result of the reduced curvature mismatch at the poles, energetically
stabilizing polar clusters and suppressing lateral ones. 
Our results are in line with the experimental observation of large polar clusters, which were found
to be robust \cite{kentner06a,kentner06b,schulmeister08} and only slightly affected by attractant 
binding \cite{homma04,lamanna05}, 
expression level variation \cite{kentner06a,maddock93,skidmore00}, and receptor methylation 
\cite{lybarger99,liberman04,shiomi05,mcandrew06}.

\begin{figure}
\includegraphics[width=8cm,angle=0]{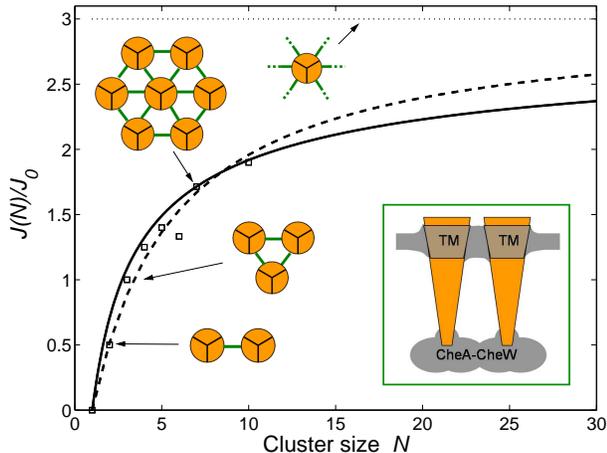}
\caption{Average interaction energy $J(N)$ per trimer (in units of coupling strength $J_0$) 
as a function of cluster size (number of trimers $N$). Trimers are shown by orange disks,
trimer-trimer coupling of strength $J_0$ is shown by green bars. 
Clusters are assumed to have a triangular lattice structure. 
Symbols: small compact clusters of minimal circumference.
Solid line and dashed lines: ``Ring model'' and ``Fit-to-symbols model'', 
respectively (see Model section). Horizontal dotted line:
average interaction energy for infinitely large cluster. Inset: Possible 
mechanisms of trimer-trimer coupling including coupling mediated by receptor-bound
CheA-CheW and membrane deformations based on large hydrophobic trans-membrane 
domains (TMs).}
\label{fig:fig3}
\end{figure}

\begin{figure}          
\includegraphics[width=8cm,angle=0]{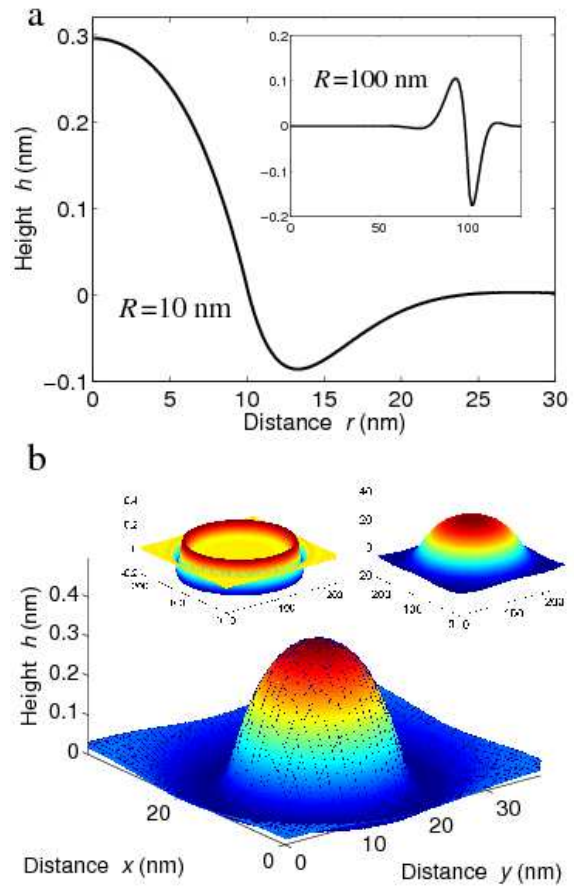}
\caption{Cluster-membrane height profiles.
(a) Profile $h(r)$ as a function of distance $r$ from center for a small cluster 
of radius $R=10$ nm. Inset: Profile for large cluster of radius $R=100$ nm. 
(b) Three-dimensional profile for small cluster. Left inset: Same
for large cluster. For parameters $\kappa_c$, $\kappa_m$, and $\lambda$ see table I.   
Right inset: large cluster for very small pinning modulus ($\lambda=10^{-5} k_BT/nm^4$).}
\label{fig:fig4}
\end{figure}

Fig. 4 shows typical cluster-membrane height profiles for two different cluster radii, $R$=10 and 100 nms. 
The height profile minimizes the cluster-membrane elastic energy given in Eq. \ref{eq:Eel}. 
The profile of the small cluster bulges out into the periplasmic space in a convex manner, 
while the large cluster of physiological size is flattened. The latter effect
appears consistent with images from cryo-electron microscopy \cite{zhang07,briegel08}. 
In our model, large deformations are suppressed by the pinning potential (left inset in Fig. 4b). 
Significant reduction of the pinning modulus $\lambda$ leads to strongly curved clusters (right inset in Fig. 4b). 
Note that the maximal deformation of the large curved cluster can easily exceed 
the width of the periplasmic space (20 nm \cite{matias03}), 
emphasizing the importance of the pinning modulus.

How large are clusters and how do their sizes differ between cell poles and lateral positions?
Consider a membrane-embedded cluster of radius $R$ or number of trimers $N\approx R^2\pi/A$, 
where $A$ is the trimer cross section. The total trimer energy $E$ is equal to the sum of the 
unfavorable elastic energy $E_{el}$ and the favorable attractive energy $E_a$. Dividing by the 
number of trimers $N$ results in the corresponding energy densities $\epsilon$, $\epsilon_{el}$, 
and $\epsilon_a$. Generally, the minimum of the total energy density $\epsilon$ as a function of cluster size
$N$ provides the preferred cluster size. Fig. 5a shows that key requirements for stable polar clusters are fulfilled: 
(1) The energy density reaches its lowest value in the limit $N\rightarrow\infty$, 
energetically favoring maximal clusters. (2) Although the energy-density difference $\Delta\epsilon$ 
between the poles and lateral cell area can be smaller than the thermal energy $k_BT$, a cluster 
of $N$ trimers is stabilized at the poles and suppressed at the lateral area when
$N\Delta\epsilon$ is larger than $k_BT$ (see Model section).

Due to the finite trimer density, cluster sizes are always finite. To obtain the predicted distribution 
of cluster sizes, we consider the combined system of cell poles and lateral area.
The distribution of cluster sizes can be calculated using Boltzmann statistics of 
equilibrium statistical mechanics (see Model section). A chemical potential is further adjusted to 
obtain a certain target trimer density. Fig. 5b shows the size distributions of polar and 
lateral clusters for three different trimer densities. Even at low trimer densities, 
very few residual trimers, stabilized by entropy, remain unclustered at the poles and lateral area.
The average radii of polar clusters shown in Fig. 5c correspond well with the observed cluster diameters 
of about $200$ nm, whereas lateral clusters are significantly suppressed.

\begin{figure}
\includegraphics[width=8cm,angle=0]{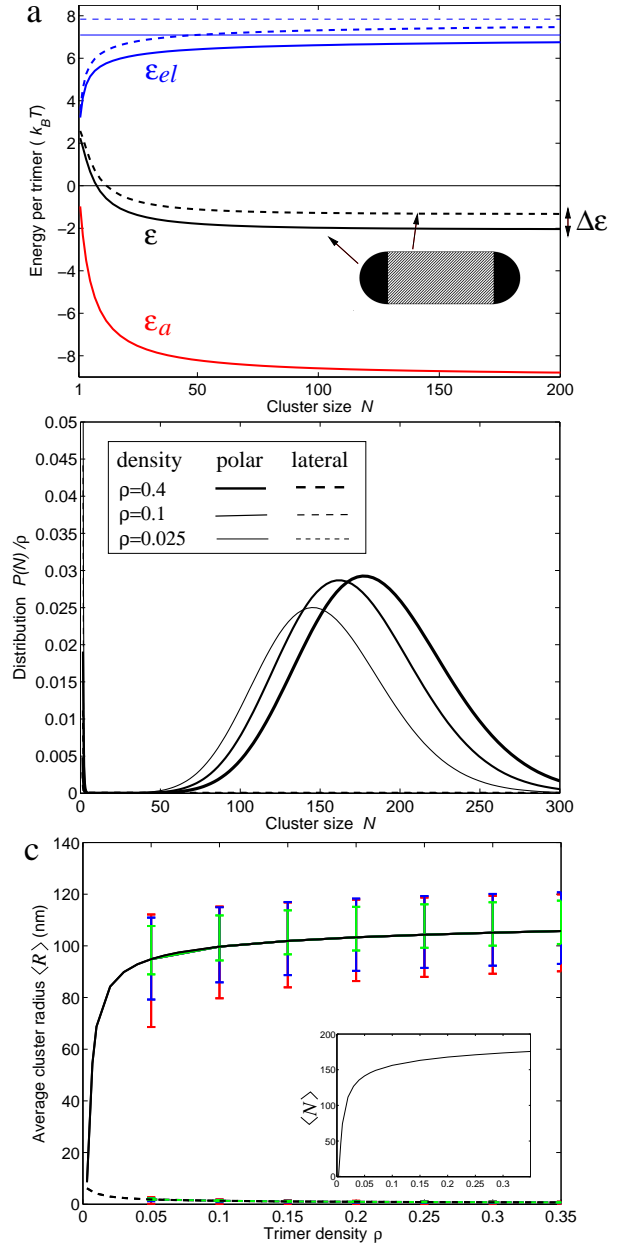}
\caption{Quantifying polar receptor clustering. (a) Cluster-membrane energy density, 
{\it i.e.} energy per trimer, as function of cluster size $N$.
Total energy density $\epsilon$ (black) is sum of repulsive elastic energy density
$\epsilon_{el}$ (blue) and attractive cluster energy density $\epsilon_a$ (red). 
Solid lines correspond to cell poles, dashed lines to lateral positions.
Thin solid and dashed blue lines indicate asymptotic limit of the elastic energy density for 
$N\rightarrow\infty$ at the poles and lateral positions, respectively. 
Polar clusters are stabilized by energy density $\Delta\epsilon$. 
(b) Distribution of cluster sizes for three values of the trimer density (occupancy fraction) 
$\rho=0.4, 0.1,$ and $0.025$ at the poles (solid lines) and lateral positions (dashed lines).
(c) Average cluster radius $\langle R\rangle$ at the poles (black solid line) and lateral area 
(black dashed line) as a function of trimer density $\rho$, based on
standard parameters, including trimer-trimer coupling $J_0$, trimer curvature $C_T$, and 
cluster bending stiffness $\kappa_{c}$ from table I.
Error bars indicate robustness to parameter changes. Upper error bars for poles and lateral area:
1.1 $J_0$ (red), 0.98 $C_T$ (dimer-dimer distance reduced by 1 nm, green), 0.92 $\kappa_c$ (blue);
lower error bars for poles and lateral area: 
0.9 $J_0$ (red), 1.03 $C_T$ (dimer-dimer distance increased by 1 nm, green), 1.09 $\kappa_c$ (blue).
Inset: Average cluster size (number of trimers) $\langle N\rangle$ as a function of trimer density
for standard parameters.}
\label{fig:fig5}
\end{figure}

As shown in Fig. 5c, our model predicts the average cluster size as function of trimer density. 
This prediction can experimentally be tested through imaging.
The fluorescence intensity of a cluster, {\it e.g.}  measured using receptor-GFP fusion proteins \cite{shiomi06}, 
is proportional to the number of receptors in the cluster.  
Alternatively, imaging by cryo-electron microscopy can provide spatial cluster dimensions \cite{zhang07,briegel08}. 
The dependence on trimer density can be studied by expressing receptors from an inducible plasmid. 
An recent experiment using the fusion protein CheY-YFP
as a fluorescent marker indeed indicated a strong correlation between receptor expression level
and polar fluorescence intensity \cite{thiem08}.
Furthermore, the predicted link between membrane curvature and clustering can be tested
by quantifying receptor-fluorescence intensities for cell-shape phenotypes, {\it e.g.} when 
inhibiting actin-homologue MreB \cite{shiomi06}, responsible for rod shape in bacteria.
Alternatively, cocci cells or round membrane vesicles can be used, allowing the study of
receptor clustering in presence of only a single membrane curvature. In Fig. 6, we show the
predicted average cluster size as a function of coccus radius and receptor density 
(expression level). Increasing the coccus radius decreases the coccus curvature and
leads to a larger cluster-membrane curvature mismatch, which reduces cluster size.
In contrast, increasing the receptor density shifts cluster size distribution toward larger clusters. 

In recent experiments the physical response of dimers were measured 
by homo-FRET using receptor-YFP fusions \cite{vaknin06,vaknin07}.
These data indicate that the dimer-dimer distance in a trimer (distance between 
receptor C-termini) shrinks by 10\% upon osmolyte stimulation. 
Osmolytes act as repellents and are presumably sensed through receptor-membrane 
coupling. To see if polar clustering is robust against such 
perturbations, Fig. 5c shows that a 10\% increase of the dimer-dimer distance 
($C_T^{-1}=36.2$ nm, lower green error bars) destabilizes clusters very little, 
wheres a 10\% decrease of the dimer-dimer distance ($C_T^{-1}=38.8$ nm, upper green error bars) 
stabilizes clusters even further.  To illustrate the robustness of polar clustering with respect 
to other model parameters, we also varied cluster bending stiffness $\kappa_c$ and trimer-trimer
coupling strength $J_0$. Fig. 5c shows that these parameter variations only lead to
modest changes in cluster stability (blue and red error bars, respectively).

\begin{figure}             
\includegraphics[width=8.0cm,angle=0]{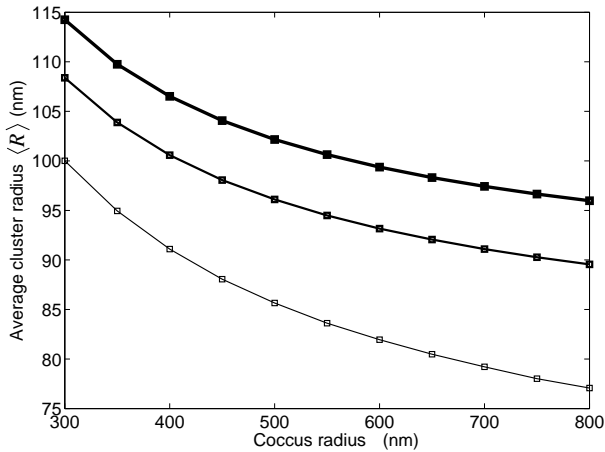}
\caption{Predicted receptor clustering in cocci cells. 
Average cluster radius $\langle R\rangle$ as a function of coccus radius.
Solid lines of varying thickness correspond to the three densities from Fig. 5b.}
\label{fig:fig6}
\end{figure}

While our model provides a robust mechanism for the formation of large polar clusters, 
there are limitations of the model. (1) A recent study showed that, in addition to
large polar clusters, there are also small lateral clusters at future division sites, 
such as 1/2, as well as 1/4 and 3/4 cell length \cite{thiem07}. 
However, these lateral clusters appear immobile, presumably due to anchoring, 
and hence may form through a different mechanism. In {\it Rhodobacter sphaeroides} immobile 
lateral clusters of chemotaxis receptor homologues were even found in the cytoplasm \cite{thompson06}.
Our model does suggest cluster formation at the new poles once cell division occurs and 
the membrane pinches off. Newly synthesized receptors, inserted by the Sec-machinery throughout the cell surface
\cite{gebert88,shiomi06}, would begin to cluster at the new cell poles. 
However, if equilibration is too slow to grow new polar clusters from scratch after cell division,
lateral clusters may be useful by serving as nucleation sites.
(2) Structural work on receptors and receptor-bound proteins in {\it Thermotoga maritima} suggests that, 
at least for this bacterium, receptor dimers may assemble into linear oligomers and 
not trimer-based clusters \cite{park06}. 
A bioinformatics study on chemotaxis receptors across many species also addresses this issue, but 
does not favor one model over the other \cite{zhulin07}. On the other hand, cryo-electron microscopy
of {\it Caulobacter crescentus} strongly supports trimer-based clusters \cite{briegel08}.
(3) Our model does not include interactions between multiple clusters. Such interactions may largely be
unimportant, since fluorescence images generally indicate only one, rarely more clusters per 
cell pole \cite{thiem07}. (4) It can further not be ruled out that
receptors do not localize to the poles themselves, but that certain polar lipids, {\it e.g.} cardiolipin
\cite{dowhan00}, provide favorable sites for receptor localization and clustering. 
(5) Although some elastic properties are included in our model, others are not, {\it e.g.} membrane
thickness deformation due to large hydrophobic transmembrane domains \cite{wiggins05}. 
Thickness deformation leads to a line tension, {\it i.e.} an elastic energy proportional to the
cluster circumference $2\pi R$, affecting both polar, as well as lateral clusters equally.
While this effect does not change the stability of polar clusters, the line tension may provide a 
mechanism for trimer-trimer coupling \cite{dan93} (see next paragraph) and have influence on
the cluster size distribution.

What are the possible mechanisms responsible for trimer-trimer coupling? 
(1) Coupling mediated by CheA and CheW (Fig. 3, inset). Presence of CheA and CheW increases
polar clustering only modestly \cite{maddock93,sourjik00,liberman04}. However, overexpression
of CheA decreases, whereas overexpression of CheW increases the receptor cooperativity 
measured by FRET \cite{sourjik04}. (2) Coupling mediated by elastic membrane deformations. 
Receptor activity has been shown to depend on receptor-membrane interactions \cite{draheim05,vaknin06,vaknin07},
which, we speculate, may provide a mechanism for trimer-trimer coupling \cite{dan93,aranda96}.
In support of this mechanism, receptor transmembrane regions are unusually large (24 to 30 residues) 
compared to the membrane thickness ($30 \AA$ or about 20 residues) \cite{boldog04}. Such large
regions possibly lead to significant membrane deformations to protect the hydrophobic
receptor residues from water (Fig. 3, inset). 
If transmembrane regions change with activity, {\it e.g.} within the
receptor piston model \cite{ottemann99,peach02,miller04,draheim05}, 
trimer-trimer coupling may even depend on the receptor
activity state, as proposed for the approximate two-state osmolarity-sensing MscL pore \cite{ursell07}. 
(3) Coupling mediated by swapping of cytoplasmic domains of neighboring dimers \cite{wolanin04}.

Many other sensory receptors cluster as well, including B-cell \cite{schamel00}, T-cell \cite{germain99},
Fc$\gamma$ \cite{chacko94}, synaptic \cite{griffith04}, and ryanodine \cite{yin05} receptors.
This indicates that receptor clustering is an important regulatory mechanism of the cell, 
{\it e.g.} to adjust signaling properties, recruit auxiliary proteins, or kinetically proof-read 
unexpected stimuli. Unlike bacteria, eukaryotic receptor clustering appears much more dynamic, 
including receptor diffusion and internalization, and the underlying physical mechanism remain 
little understood. Our work adds additional support to the idea that the 
elastic properties of receptors and membrane may be a general design principle 
to regulate receptor localization and clustering \cite{brown08,groves07}.

\appendix

\section{Surface tension and Gaussian curvature}
The elastic energy in Eq. \ref{eq:Eel} neglects surface tension
and Gaussian curvature terms. Here we show that these two elastic energy 
contributions are indeed very small.\\

\noindent{\bf Surface tension}\\ In the small deformation approximation 
(Monge representation) for cluster and membrane, 
the contribution from the surface tension to the elastic energy Eq. \ref{eq:Eel} is given by 
\begin{equation}
E_{el}^{\sigma}=\frac{\sigma_c}{2}\int_{c}(\vec\nabla h_c)^2d^2\vec r+\frac{\sigma_m}{2}\int_{m}(\vec\nabla h_m)^2d^2\vec r\label{eq:Eelo}
\end{equation}
where $\sigma_{c(m)}$ is the surface tension of the cluster (membrane) and
$\vec\nabla=(\partial/\partial x,\partial/\partial y)$.
Surface tension arises in part from the attractive receptor-receptor and lipid-lipid interactions.
Minimization of the total elastic energy (Eq. \ref{eq:Eel} plus Eq. \ref{eq:Eelo})
leads to the Euler-Lagrange equation
\begin{equation}
\nabla^4h_{c(m)}(\vec r)-\frac{\sigma_{c(m)}}{\kappa_{c(m)}}\nabla^2h_{c(m)}(\vec r)+
\frac{\lambda}{\kappa_{c(m)}} h_{c(m)}(\vec r)=0\label{eq:ELo}
\end{equation}
for the cluster (membrane).
Using integration-by-parts twice \cite{landau64,wiggins05} and Eq. \ref{eq:ELo}, the elastic energy 
contribution due to the surface tensions of the cluster and membrane is given by
\begin{equation}
E_{el}^\sigma=\pi(\sigma_c-\sigma_m)RH_1S\label{eq:Eelo2},
\end{equation}
where $R$ is the cluster radius, $H_1$ is the cluster-membrane height at the interface $(r=R)$, and
$S$ is the slope of the cluster-membrane at the interface.
This energy describes a line tension ($\sim2\pi R$) and vanishes for $\sigma_c=\sigma_m$ since
cluster and membrane contributions point in opposite radial directions.

We apply perturbation theory to estimate the significance of the surface-tension
contribution. For this purpose, we
use our previously calculated height profile, Eqs. \ref{eq:solc} and \ref{eq:solm}, obtained
without the surface tension term. Using $\sigma_m=\sigma_c/4=1\,k_BT/\text{nm}^2$ \cite{wiggins05} and 
a physiological cluster radius $R=100$ nm, we find that the estimated energy contribution from
the surface tensions is much smaller than the elastic energy Eq. \ref{eq:Eel}, 
{\it i.e.} $E_{el}^\sigma/E_{el}\approx 0.004$, justifying the neglect of this term.\\ 

\noindent{\bf Gaussian curvature}\\ The contribution to the elastic energy from the Gaussian curvature
can be neglected for homogeneous membranes that do not change their topology (Gauss-Bonnet theorem).
However, this contribution is technically non-zero for our cluster-membrane system.
In the small deformation approximation, this contribution is given by \cite{boal02} 
\begin{eqnarray}
E_{el}^G&=&K_{G,c}\int_{c}\left[\frac{\partial^2h_c}{\partial x^2}\,\frac{\partial^2h_c}{\partial y^2}
-\left(\frac{\partial^2h_c}{\partial x\partial y}\right)^2\right]d^2\vec r\nonumber\\
&&\!\!\!\!\!\!\!\!\!\!\!\!\!\!\!\!\!\!+K_{G,m}\int_{m}\left[\frac{\partial^2h_m}{\partial x^2}\,\frac{\partial^2h_m}{\partial y^2}
-\left(\frac{\partial^2h_m}{\partial x\partial y}\right)^2\right]d^2\vec r,\label{eq:EelG}
\end{eqnarray}
where $K_{G,c(m)}$ is the Gaussian curvature modulus for the cluster (membrane). 
In \cite{wiggins05}, Eq. 129 shows that the Gaussian energy contribution has two
parts. One topological part, which is just a constant since our
membrane contains a single receptor cluster.  
The second part is a contour integral along the cluster boundary. Based on Eq. 131 
of \cite{wiggins05}, we obtain
\begin{equation}
E_{el}^G=\pi(K_{G,c}-K_{G,m})S^2\label{eq:EelG}.
\end{equation}
For our height profile Eq. \ref{eq:solc} and \ref{eq:solm}, as well as $K_{G,c(m)}=-\kappa_{c(m)}/2$ \cite{boal02,wiggins05}, 
this energy contribution is significantly smaller than the elastic energy in Eq. \ref{eq:Eel}, 
{\it i.e.} $|E_{el}^G|/E_{el}\approx 0.0006$, justifying the neglect of this term. 
As expected, this energy contribution vanishes when the elastic properties of
cluster and membrane become identical for $K_{G,c}=K_{G,m}$.
\\

\begin{acknowledgments}
We thank Kerwyn Huang, Michael Manson, Ranjan Mukhopadhyay, Samuel Safran, Victor Sourjik, Sriram Subramaniam, and
Ned Wingreen for helpful discussions and two anonymous referees for valuable suggestions. 
We further acknowledge funding from the BBSRC grant BB/G000131/1 and from the
Centre for Integrated Systems Biology at Imperial College (CISBIC).
\end{acknowledgments}

\end{document}